\documentclass[fleqn,usenatbib]{mnras}

\usepackage{graphicx}
\usepackage{txfonts}
\usepackage{natbib}
\bibpunct{(}{)}{;}{a}{}{,}
\usepackage{url}
\usepackage{hyperref}
\usepackage{pdflscape}

\hypersetup{
  colorlinks=true,   
  urlcolor=blue,     
  linkcolor=blue,   
  citecolor=blue
}

\let\textcite\citet
\let\cite\citep 

   \title[The convective kissing instability]{The convective kissing instability in low-mass M-dwarf models: convective overshooting, semi-convection, luminosity functions, surface abundances and star cluster age dating}

\author[S. Mansfield \& P. Kroupa]{
Santana Mansfield$^{1}$,
         Pavel Kroupa$^{1,2}$
\\
$^{1}$Helmholtz-Institut f\"ur Strahlen- und Kernphysik (HISKP), Universit\"at Bonn, Nu\ss allee 14-16, D-53115 Bonn, Germany\\
$^{2}$Astronomical Institute, Faculty of Mathematics and Physics, Charles University in Prague, V  Hole\v{s}ovi\v{c}k\'ach 2, CZ-18000 Praha, Czech Republic\\
    }

\date{Accepted XXX. Received YYY; in original form ZZZ}

\pubyear{2023}

\begin{document}
\label{firstpage}
\pagerange{\pageref{firstpage}--\pageref{lastpage}}
\maketitle

\begin{abstract}
Low-mass models of M-dwarfs that undergo the convective kissing instability fluctuate in luminosity and temperature resulting in a gap in the main sequence that is observed in the $Gaia$ data. During this instability, the models have repeated periods of full convection where the material is mixed throughout the model. Stellar evolution models are performed using MESA with varying amounts of convective overshooting and semi-convection. We find that the amplitude and intensity of the instability is reduced with increasing amounts of overshooting but sustained when semi-convection is present. This is reflected in the loops in the evolutionary tracks in the Hertzsprung-Russell diagram. The surface abundances of $^1$H, $^3$He, $^4$He, $^{12}$C, $^{14}$N and $^{16}$O increase or decrease over time due to the convective boundary, however the relative abundance changes are very small and not likely observable. The mass and magnitude values from the models are assigned to a synthetic population of stars from the mass-magnitude relation to create colour-magnitude diagrams, which reproduce the M-dwarf gap as a large indent into the blueward edge of the main sequence (MS). This is featured in the luminosity function as a small peak and dip. The width of the MS decreases over time along with the difference in width between the MS at masses higher and lower than the instability. The parallel offset and relative angle between the upper and lower parts of the MS also change with time along with the mass-magnitude relation. Potential age-dating methods for single stars and stellar populations are described.

  \end{abstract}


%

\section{Introduction}

The M-dwarf gap is a recently discovered deficiency of stars at the lower end of the main sequence (MS) in the \textit{Gaia} Data Release 2 (DR2) data \citep{jao}, and the early \textit{Gaia} Data Release 3 (eDR3) dataset \citep{dr3}. The drop in density of stars at this location is understood to be due to the stars undergoing the convective kissing instability at the boundary between less-massive stars that are fully convective and slightly higher-mass ones with radiative cores and convective envelopes \cite{vansaders,baraffe,feiden,mansfield,Boudreaux_2023}. 

\begin{figure}
    \centering
    \includegraphics[width=\columnwidth]{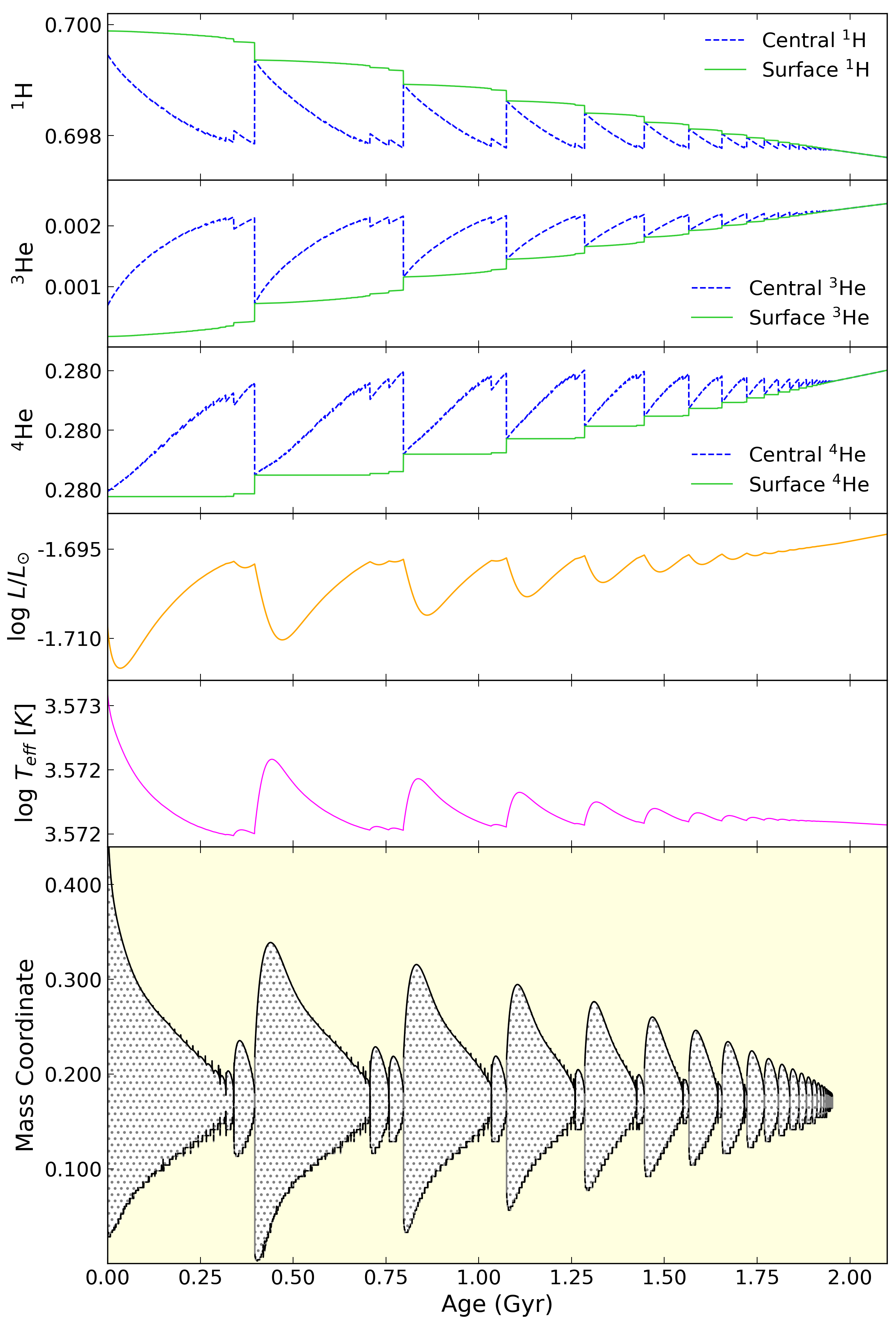}
    \caption{The top five plots give the surface and central abundances by mass fraction of hydrogen and helium, luminosity and effective temperature over time for a 0.37 $M_{\odot}$, $Z = 0.02$ model without overshooting or semi-convection}. The bottom plot shows the convective (solid yellow) and radiative (dotted grey) regions for the same model in which the convective kissing instability can be seen. 
    \label{0.37M}
\end{figure}

The phenomenon is a consequence of a delicate balance between temperature and luminosity in the criteria for convection which leads to a convective region growing at the centre of a star, and the unequal abundance of hydrogen and helium throughout the star due to the radiative region between the convective core region and convective envelope (illustrated in Fig. \ref{0.37M}). According to the Schwarzschild criterion \citep{Schwarzschild}, convection occurs in the stellar material when the radiative temperature gradient is larger than the adiabatic gradient, $\nabla_{rad} > \nabla_{ad}$, where

\begin{equation}
    \nabla_{rad} \propto \frac{L\kappa}{T^4}. 
    \label{eq4}
\end{equation}

\noindent Simplified, if the term on the right side of equation (\ref{eq4}) is small then the material is radiative, and if it is large then it is convective. 
Within the small mass range at which stars undergo the convective kissing instability, the luminosity and temperature compete to dominate equation (\ref{eq4}) such that there is a persistent unbalancing of the equation, leading to the material at the very centre of the star repeatedly switching between a convective and radiative state. As described in \citet{mansfield}, the pp-I chain is the primary thermonuclear process for M-dwarf stars

\vspace*{-2pt}
\begin{equation}
 \hspace*{19pt} p + p \longrightarrow d + e^+ + \nu_e, \ \ \ \  \label{eq1}
\end{equation}

\vspace*{-21.5pt}
\begin{equation}
 p + d \longrightarrow\ ^3\!He + \gamma, \label{eq2}
\end{equation}

\vspace*{-21.5pt}
\begin{equation}
 \hspace*{-15pt} ^3He +\ ^3\!He \longrightarrow\ ^4\!He + 2p,  \label{eq3} 
\end{equation}

\noindent where equation (\ref{eq2}) dominates. At around $0.35M_{\odot}$, a star begins on the zero-age main sequence (ZAMS) with sufficient central temperature (equation \ref{eq4}) to have a radiative core. Then with the onset of nuclear burning, which primarily produces $^3$He in equation (\ref{eq2}), enough luminosity is produced in the core that a convective region forms within the radiative one (equation \ref{eq4}). The abundance of $^3$He rises in the convective core but is unable to reach equilibrium abundance with the rest of the star due to the radiative region. As nuclear production proceeds, greater amounts of the material in the core region becomes convective until it finally merges with the convective envelope. The star undergoes a short ($\approx$ 1 Myr) period of full convection, resulting in a mixing of the stellar material. As $^3$He and $^4$He are pulled out of the core, the pp-I chain is disrupted and the nuclear productivity, and thus luminosity, subsides. A reduction in nuclear generation means the star contracts and the temperature increases. With the decrease in luminosity and rise in temperature, the material in the central region becomes radiative again according to equation (\ref{eq4}). Once the central temperature becomes high enough to produce sufficient $^3$He, which is now no longer being transported out towards the surface due to convection, equation (\ref{eq3}) can resume and the pp-I chain is restored. Then the luminosity is sufficient again to once more begin growing an inner convective region as per equation (\ref{eq4}). The fluctuations in luminosity and temperature result in loops in the Hertzsprung--Russell (HR) diagram of stellar evolution models \citep{mansfield}, and is understood to produce the M-dwarf gap. 

The process of the convective kissing instability occurs repeatedly over time but is also dampened with time due to the equilibrium abundances. Each time there is a merging of core and envelope, the helium abundance drops in the centre and rises at the surface, as helium becomes equally distributed throughout the star. When the nuclear production resumes, the central helium abundance rises but only to a certain 'critical' amount before falling again at the next merging event. At this time, because the surface abundance of helium is higher than that of the previous merging event, the decrease of central helium abundance is less before it reaches equilibrium. And so with each moment of full convection, the time it takes to reach equilibrium abundance becomes increasingly less, and correspondingly the time at which the nuclear production is diminished is lower. Accordingly then, the amount by which the luminosity drops is also reduced, resulting in a smaller region of the stellar material becoming radiative. This is what we see in Fig. \ref{0.37M} with the radiative region becoming smaller over time. Once the surface abundance reaches the critical central abundance, true equilibrium is attained and the pp-I chain can finally run without disruption. Now that luminosity is steadily produced, it is high enough (equation \ref{eq4}) to retain the stellar material in a prevailing fully convective state.

Whilst it is now understood that the convective kissing instability is a possible cause of the M-dwarf gap, some questions about the nature of the instability remain, namely how certain convective and mixing processes may alter and affect the intensity and duration of the convective kissing instability due to the blurring of boundaries between convective and radiative regions and the added mixing of material towards equilibrium abundance. These processes are primarily convective overshooting and semi-convection. The effects of these processes will be investigated in this publication along with the consequences of the instability on the surface abundances and the stellar luminosity function. 
  
  \begin{figure*}
      \centering
      \includegraphics[scale=0.25]{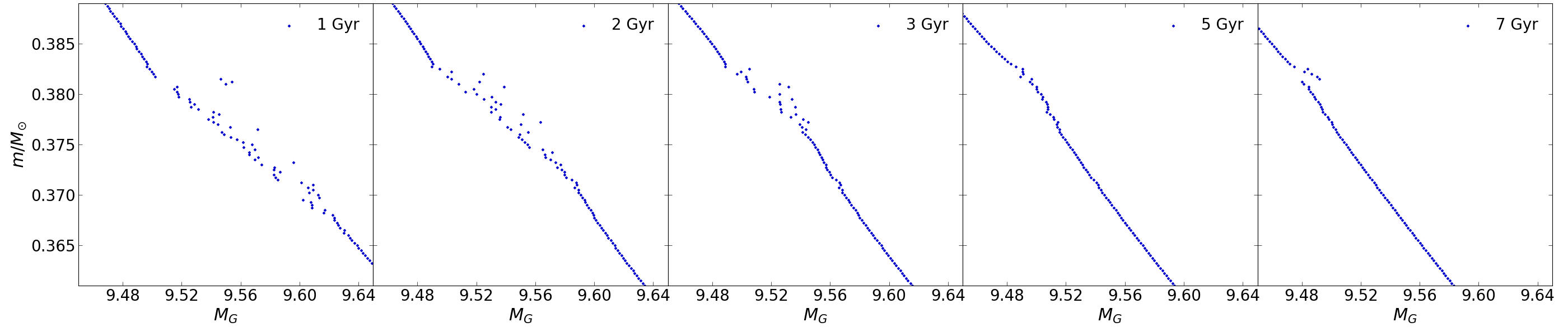}
      \caption{Mass--magnitude relation in the photometric $G-$band for models with $Z = 0.02$, without overshooting or semi-convection, at five time intervals of 1, 2, 3, 5 and 7 Gyr. The fluctuations in magnitude caused by the convective kissing instability create a discontinuity in the relation.}
      \label{mass-mag}
  \end{figure*}

Convective overshooting occurs when the material in a convective cell has a non-zero velocity when it reaches the boundary of the cell and so continues to move a small distance past it, mixing the material into the radiative region. The extent of overshooting is mass dependent \citep{claret} and so for low-mass stars only a small amount of overshooting will be present. Yet even at these small numbers, the ability for material to be mixed across the boundary will alter the distribution of helium abundance and likewise the amplitude of the convective kissing instability.

Semi-convection is a diffusive process in a material that occurs in a region that is unstable to convection as per the Schwarzschild criterion, but stable according to the Ledoux criterion \citep{ledoux}. This happens when:

\begin{equation}
\nabla_{ad} < \nabla_{rad} < \nabla_{L},
    \label{eq_ledoux1}
\end{equation}

\noindent with the Ledoux temperature gradient:

\begin{equation}
\nabla_L = \nabla_{ad} + \frac{\varphi}{\delta}\nabla_{\mu},
    \label{eq_ledoux2}
\end{equation}

\noindent and

\begin{equation}
\varphi = \left(\frac{\delta\  \textnormal{ln}\ \rho}{\delta\  \textnormal{ln}\ \mu} \right)_{P,T}, \delta = \left(\frac{\delta\  \textnormal{ln}\ \rho}{\delta\  \textnormal{ln}\ T} \right)_{P,\mu}, 
\nabla_{\mu} = \left(\frac{\delta\  \textnormal{ln}\ \mu}{\delta\  \textnormal{ln}\ P} \right),
    \label{eq_ledoux3}
\end{equation}

\noindent where $\rho$ is the density, $\mu$ is the mean molecular weight, $P$ is the pressure and $T$ is the temperature. When a non-homogeneous composition of material remains after a convective region shrinks, a discontinuity of density and molecular weight results in a slow mixing of material called semi-convection. This process would happen in M-dwarf stars undergoing the convective kissing instability due to convective regions repeatedly growing and shrinking.

\begin{figure*}
    \includegraphics[scale = 0.1]{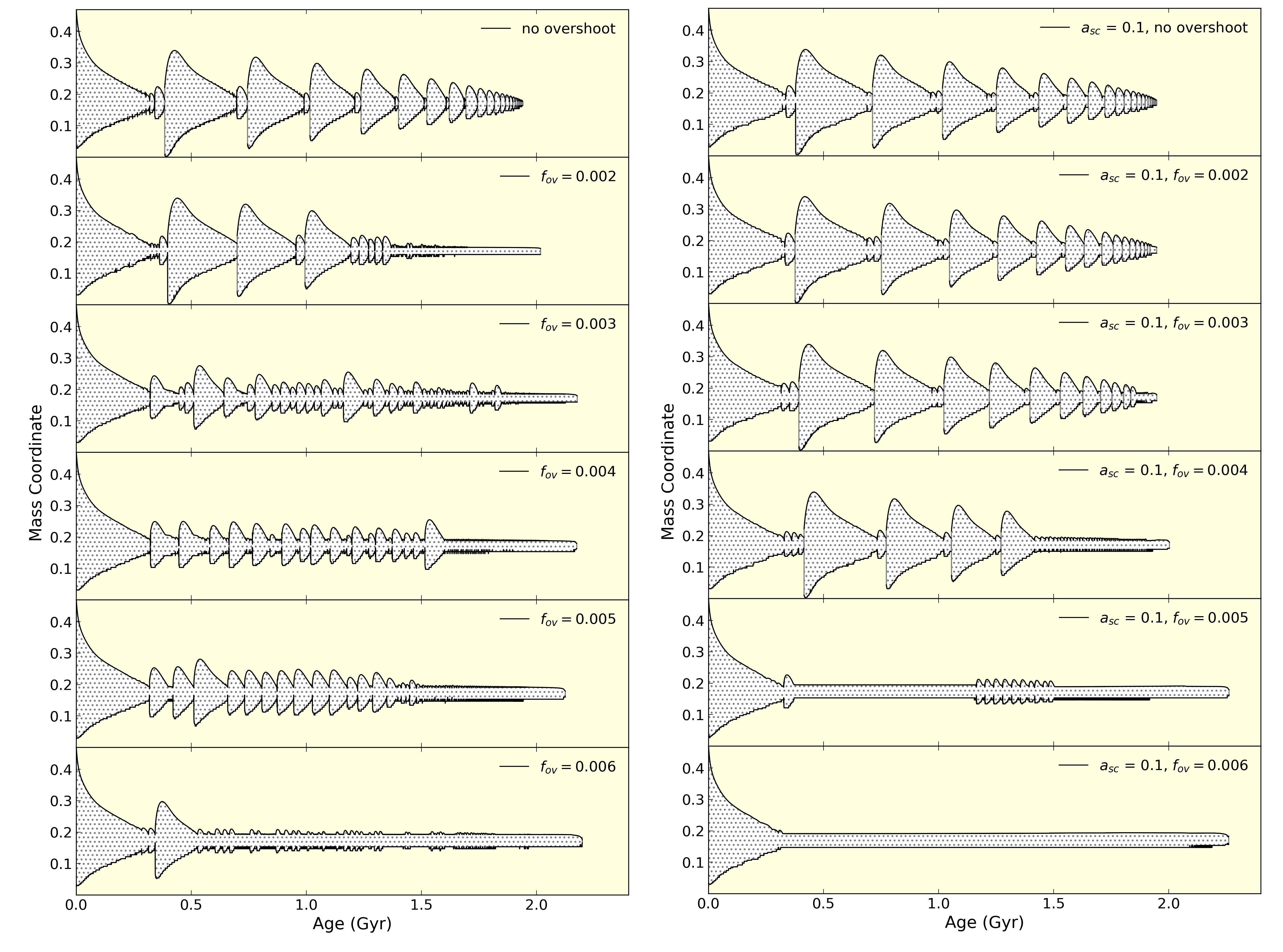}
    \caption{Mass coordinate of the convective (yellow solid) and radiative (grey dotted) regions plotted over time for a 0.37$M_{\odot}$, $Z = 0.02$ model. The models on the left used the Schwarzschild criterion for convection with no semi-convection but increasing overshooting parameter $f_{ov}$, and the models on the right used the Ledoux criterion with semi-convection parameter $\alpha_{sc} = 0.1$ and also increasing $f_{ov}$.}
    \label{overshoot_semi}
\end{figure*}

This work will also look at the changes in surface abundances of the models, as the abundance of $^3$He is seen to increase each time the core and envelope merges \cite{mansfield}. Until now only the effects of the instability on $^3$He have been previously investigated. We will look to see if the convective kissing instability influences the surface abundances of other light elements, particularly $^1$H, $^{12}$C and $^{16}$O as these affect the spectra of M-dwarfs. The cool temperatures of M-dwarf atmospheres allows for the formation of molecules and dust particles, which leads to considerable molecular absorption in their spectra \citep{allard1995}. The main molecules responsible for the opacity of M-dwarf atmospheres are those which contain either C, O or H, namely TiO and VO that dominates the optical spectrum \citep{allard}, and H$_2$O and CO that prevails in the infrared \citep{veyette,Rajpurohit2013}. Changes in the surface abundances may result in alterations of the atmosphere properties. For example, a reduction of carbon and oxygen would
lead to less TiO, CO and H$_2$O forming, and thus a different configuration of the surface material for stars of lower masses. This would then affect the properties and structure of M-dwarf atmospheres, which are sensitive to these abundances. For example, \citet{allard} found that a reduction in H$_2$O and TiO in models results in a cooling of the atmosphere.

Additionally we will explore the consequence of the convective kissing instability on the stellar luminosity function (LF), which is given by:
\begin{equation}
    \phi_p = \frac{dN}{dM_p} = - \frac{dN}{dm}\frac{dm}{dM_p} = -\xi \frac{dm}{dM_p},
    \label{1}
\end{equation}

\noindent where $N$ is the number of stars per cubic parsec, $M_p$ is the absolute magnitude in a photometric pass band $p$, $m$ is the stellar mass, and $\xi (m) dm$ is the number of stars per cubic parsec with masses between $m$ and $m +dm$. Specific features are present in the stellar LF which are due to the internal material of stars at certain masses. For solar metallicity stars, the LF has a minimum at $M_V \approx 7$ called the Wielen dip \citep{wielen}, due to H$^-$ becoming a more important source of opacity for decreasing mass. Once it becomes the primary source, there is a flattening of the luminosity--mass relation and the dip in the luminosity function \cite{pavel}. The stellar LF also has a maximum at $M_V \approx 11-12$ due to the onset of H$_2$ formation in a thin outer shell \citep{pavel}. For lower metallicities, these features become shifted to brighter magnitudes \cite{pavelmasslum}. Upon discovering the M-dwarf gap,  \citet{jao} predict that there should be a dip in the LF at around $M_V \approx 10.5 - 11.5$ for solar metallicities. 

The fluctuations in luminosity during the convective kissing instability causes a discontinuity in the luminosity--mass relation \citep{mansfield}. The LF depends on the slope of the luminosity--mass relation, and as such, the discontinuity means that the relation is indifferentiable and thus the slope, $dm/dM_p$, cannot be calculated at the mass range of the convective kissing instability. In this work we will use an alternative method to determine the LF by creating a synthetic population of stars with properties that are assigned using the results from stellar evolution models. As the LF represents the number of stars per magnitude bin, we can recreate the LF from a count of the number of stars within the magnitude bins. We also consider the mass-magnitude relation and its derivative, as well as potential age-dating methods for single stars and populations such as star clusters.

This investigation uses stellar evolution models which are detailed in Sect. \ref{models}. The setup for a synthetic population of stars using the results of these models is described in Sect. \ref{pop}. The results are presented in Sect. \ref{results}, a discussion is given in Sect. \ref{discus}, followed by a summary in Sect. \ref{sum}.

\begin{figure}
    \centering
    \includegraphics[width=\columnwidth]{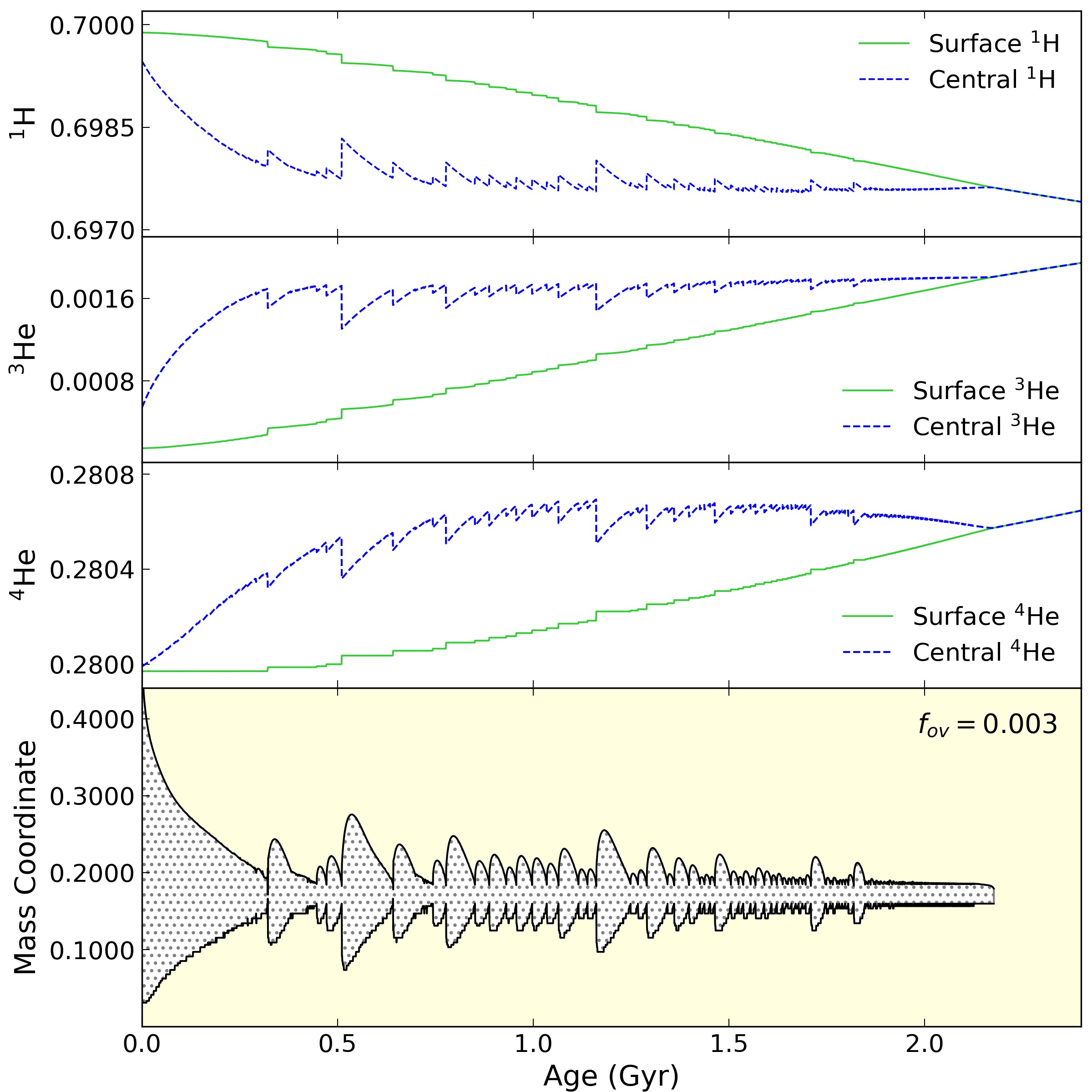}
    \caption{Central and surface abundances by mass fraction for the $0.37 M_{\odot}$, $Z = 0.02$ model with overshooting parameter $f_{ov} = 0.003$, along with the mass coordinate of the radiative (grey dotted) and convective (yellow solid) regions over time.}
    \label{3he_crit_abund}
\end{figure}

  \begin{figure}
      \centering
      \includegraphics[width=\columnwidth]{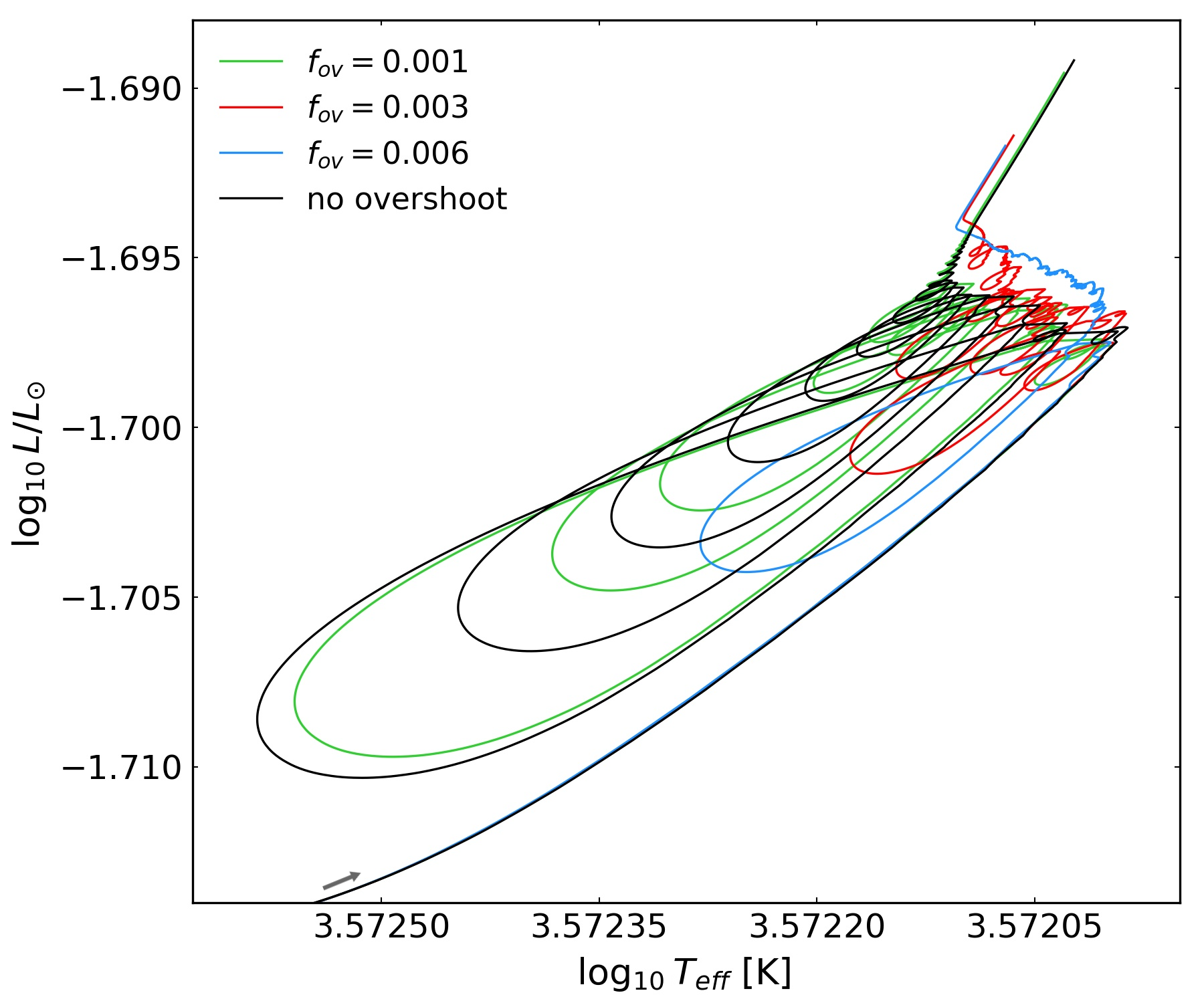}
      \includegraphics[width=\columnwidth]{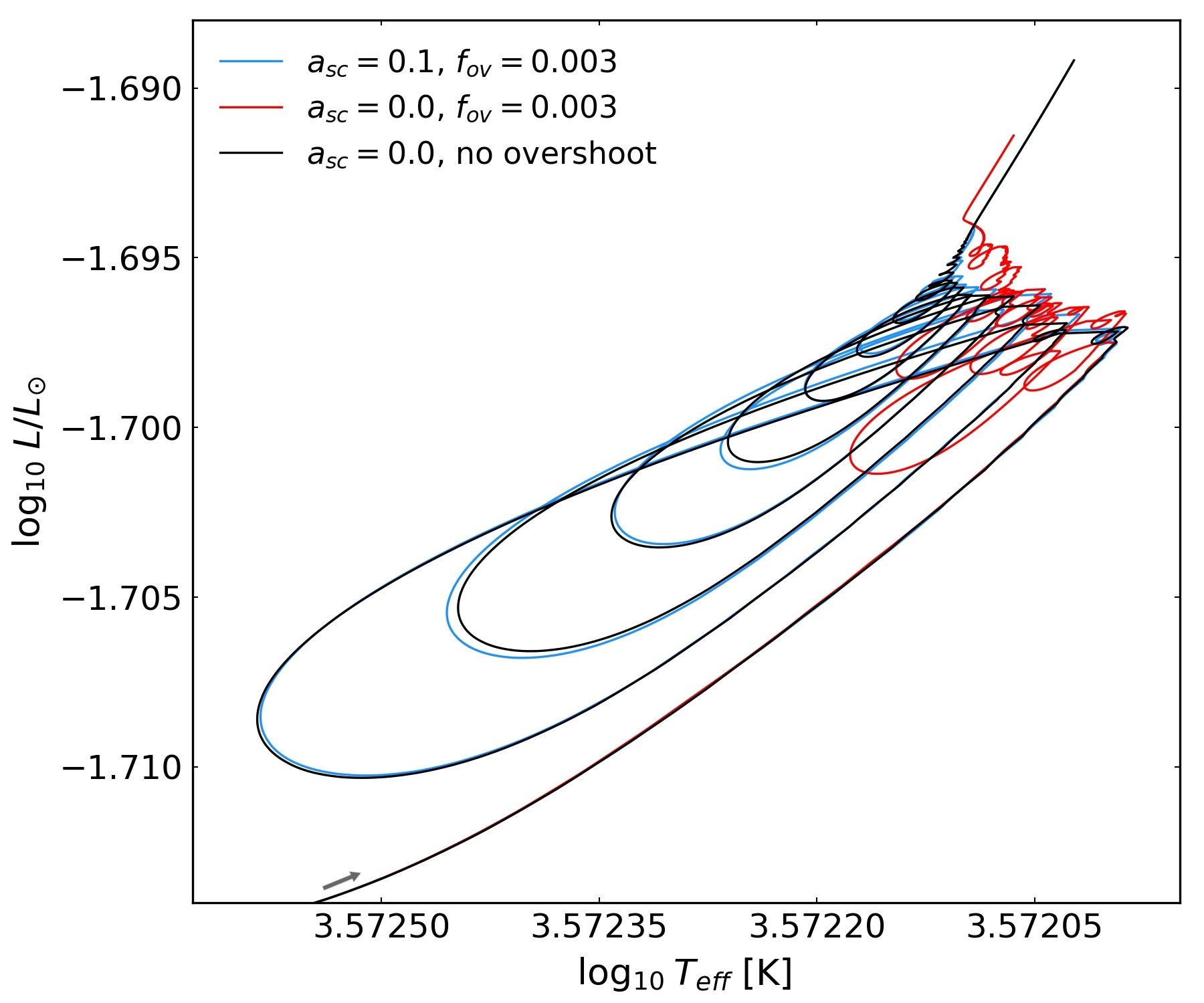}
      \caption{HR diagram showing the evolutionary track of the $0.37 M_{\odot}$, $Z = 0.02$ model with varying amounts of overshooting (\textit{top}), where increasing overshooting diminishes the loops. The models evolve from the lower left as indicated by the arrow. The bottom plot shows the tracks with no overshooting or semi-convection (black), with overshooting added (red) and with both overshooting and semi-convection (blue) which sustains the loops.}
      \label{overshoot_semi_HR}
  \end{figure}

\section{Stellar Models}\label{models}

The 1D stellar evolution code Modules for Experiments in Stellar Astrophysics
(MESA, version r22.11.1) is utilised (\citealp{Paxton2011,Paxton2013,Paxton2015,Paxton2018,Paxton2019}). The equations of state (EOSs) applied in MESA are a combination of the SCVH \cite{scvh}, HELM \cite{helm}, OPAL \cite{opal}, FreeEOS \cite{Irwin2004}, and PC \cite{Potekhin2010} EOSs. Molecules and dust grains are able to form in the low temperatures of M-dwarf atmospheres and the opacity tables for these temperatures are from \textcite{ferguson}. The nuclear reaction rates are from \textcite{Cyburt2010}, with additional weak reaction rates \cite{Fuller1985, Oda1994, Langanke2000}. The opacity tables are those given by OPAL (\citealp{Iglesias1993},
 \citealp{Iglesias1996}) and \textcite{gs}, as well as electron conduction opacities from
  \textcite{Cassisi2007}. Thermal neutrino loss rates are from \textcite{Itoh1996}. Screening was included using the prescription of \textcite{Chugunov2007}. For the outer boundary conditions of M-dwarfs, the \textcite{hauschildt} model atmosphere tables were used, taken at $\tau = 100$ \cite{Paxton2011, chabrier1997}. The initial helium abundance is Y = 0.24 + 2Z by mass fraction. Convective mixing is described using mixing length theory (mixing length parameter $\alpha_{MLT} = 2$), with the convective premixing scheme (CPM, \citealt{Paxton2019}) which iterates over each cell and instantaneously applies mixing to the cells which are found to be convective (\citealp{Ostrowski}). Sets of models are calculated at solar metallicity ($Z = 0.020$) and a lower metallicity of $Z = 0.001$ to represent a globular cluster (GC). They begin at the ZAMS and run until 9 Gyr. A maximum time step of 2 $\times 10^5$ years is used in order to slow the code down during the MS where the convective kissing instability occurs. Similarly to that described in \citet{mansfield}, this timestep in small enough to see the instability with sufficient resolution, but large enough that the computational time is reasonable. Binary systems were considered but not applied here for simplicity. The conclusions drawn should be similar for binary systems as the convective kissing instability affects the stars individually. 
  
 Convective overshooting is administered by an exponentially decaying, time-dependent diffusive processes past the convective boundary given by

 \begin{equation}
D_{ov} = D_0\ {\rm exp} \left(- \frac{2r}{f_{ov}H_p} \right),
     \label{eq_conv}
 \end{equation}

\noindent where $D_0$ is the diffusion coefficient inside the boundary, $r$ is the radial distance past the boundary, $f_{ov}$ is the free overshooting parameter and $H_p$ is the pressure scale height \cite{herwig,Paxton2011}. The amount of convective core overshooting is strongly dependent on mass for $m \lesssim 2M_{\odot}$ with $f_{ov}$ increasing sharply up to $f_{ov} \approx 0.0160$ where it stays constant for masses higher (see \citealt{claret}, Fig. 10). As such, M-dwarf stars of this mass range will have a very small amount of overshooting. We used $0.0005 \leq f_{ov} \leq 0.0060$ in this work to see the effect on the models, but note that it is likely that $f_{ov} \ll 0.0050$.
 
 For semiconvection to be implemented, the Ledoux criterion for convection is used (equations \ref{eq_ledoux1} - \ref{eq_ledoux3}). Mixing in regions satisfying equation \ref{eq_ledoux1} is via a time-dependent diffusive process with diffusion coefficient

 \begin{equation}
D_{sc} = \alpha_{sc}\ \left( \frac{\kappa_r}{6 c_p \rho}\right)\ \frac{\nabla_{rad} - \nabla_{ad}}{\nabla_L - \nabla_{rad}},
     \label{eq_semi}
 \end{equation}

 \noindent where $\kappa_r$ is the radiative conductivity and $c_p$ is the specific heat at constant pressure. The values applied in this work for the dimensionless semiconvection parameter $\alpha_{sc}$ are 0.1 and 1. 

The models do not include thermohaline mixing as it is not an important mixing process for M-dwarfs on the MS. This is because for thermohaline mixing to occur, the mean molecular weight decreases inwards, which can occur in low mass stars after the first dredge up \citep{charbonnel,cantiello}, but is not present during the early MS at which the convective kissing instability occurs.

  \section{Population synthesis}\label{pop}
  
  To construct colour-magnitude diagrams (CMDs) and the luminosity function, three sets of 100,000 stars are given masses ranging from 0.08 $M_{\odot}$ to 0.60 $M_{\odot}$ according to the canonical initial mass function (IMF), $\xi (m) \propto m^{-\alpha}$, with $\alpha = 1.3$ for $m \leq 0.50 M_{\odot}$ and $\alpha = 2.3$ for $m > 0.50 M_{\odot}$ \citep{pavel2001,pavel2013}. These are then assigned an age and corresponding magnitude values due to the mass--magnitude relation determined from the MESA models which is shown in Fig. \ref{mass-mag}, where there is a discontinuity due to the fluctuations in luminosity. The slope of the mass--magnitude relation differs at masses lower than and larger than the discontinuity, and changes over time. Thus in order to assign magnitude values to masses higher and lower than the discontinuity, the equation of the line is determined at each point in time. For the masses within the instability range, the magnitude is taken directly from the corresponding MESA models. This method is applied because the slope, $dm/dM_p$, within the instability range cannot be determined, as the relation is indifferentiable across the discontinuity.

Following \citet{Paxton2018}, the absolute magnitude of a star in a photometric pass band $p$ is determined by 

\begin{equation}
    M_p = M_{bol} - BC_p,
    \label{eq_absmag}
\end{equation}

\noindent with the absolute bolometric magnitude 

\begin{equation}
    M_{bol} = -2.5\ {\rm log}_{10} (L/L_{\odot}) + M_{bol,\odot},
    \label{eq_mbol}
\end{equation}

\noindent where $M_{bol,\odot} = 4.74$ is the absolute bolometric magnitude of the Sun. The bolometric corrections $BC_p$ are taken from a pre-processed set of tables provided within MESA that are described in \citet{Paxton2018} as computed from atmosphere models, and defined as functions of the stellar photosphere including the effective temperature $T_{eff}$ \citep{code,Houdashelt_2000}. The conversion to the \textit{Gaia} $G$-band magnitudes follows \citet{gaia_transformation}.

The first set has a metallicity of $Z = 0.020$ to represent stars in the local neighbourhood and the second set uses a metallicity of $Z = 0.001$ to represent stars in a globular cluster. Both of these sets have no overshooting or semi-convection applied. The third set is calculated using semi-convection with an efficiency of $\alpha_{sc} = 0.1$ and an overshooting parameter of $f_{ov} = 0.006$, with a metallicity of $Z = 0.020$. All other parameters of the sets remain the same.

  \section{Results}\label{results}
  
 Models with masses 0.3440$M_{\odot} - 0.3825M_{\odot}$ undergo the convective kissing instability for $Z = 0.020$.

  \subsection{Convective overshooting and semi-convection}\label{overshoot_sec}

  Figure \ref{overshoot_semi} illustrates  the effects of overshooting up to $f_{ov} = 0.006$ and semi-convection with efficiency $\alpha_{sc} = 0.1$ for a $0.37 M_{\odot}$ model. As $f_{ov}$ increases, the amplitude of the convective kissing instability is lessened, and there are fewer mergers of the core and envelope into a fully convective state. Instead, the core and envelope come close enough together that the remaining distance is small enough for material to cross the radiative region due to overshooting. As the abundance of helium is able to gradually increase throughout the model by passing through the radiative region (as seen by the steady increase of surface $^3$He abundance in Fig. \ref{3he_crit_abund}), there is no longer the dampening of the convective kissing instability over time that was seen when no overshooting is present (Fig. \ref{0.37M}). 
  The results here are shown up to $f_{ov} = 0.006$ however we again note that real M-dwarfs are likely to have a low amount of overshooting \citep{claret}.

   \begin{figure}
      \centering
      \includegraphics[scale = 0.42]{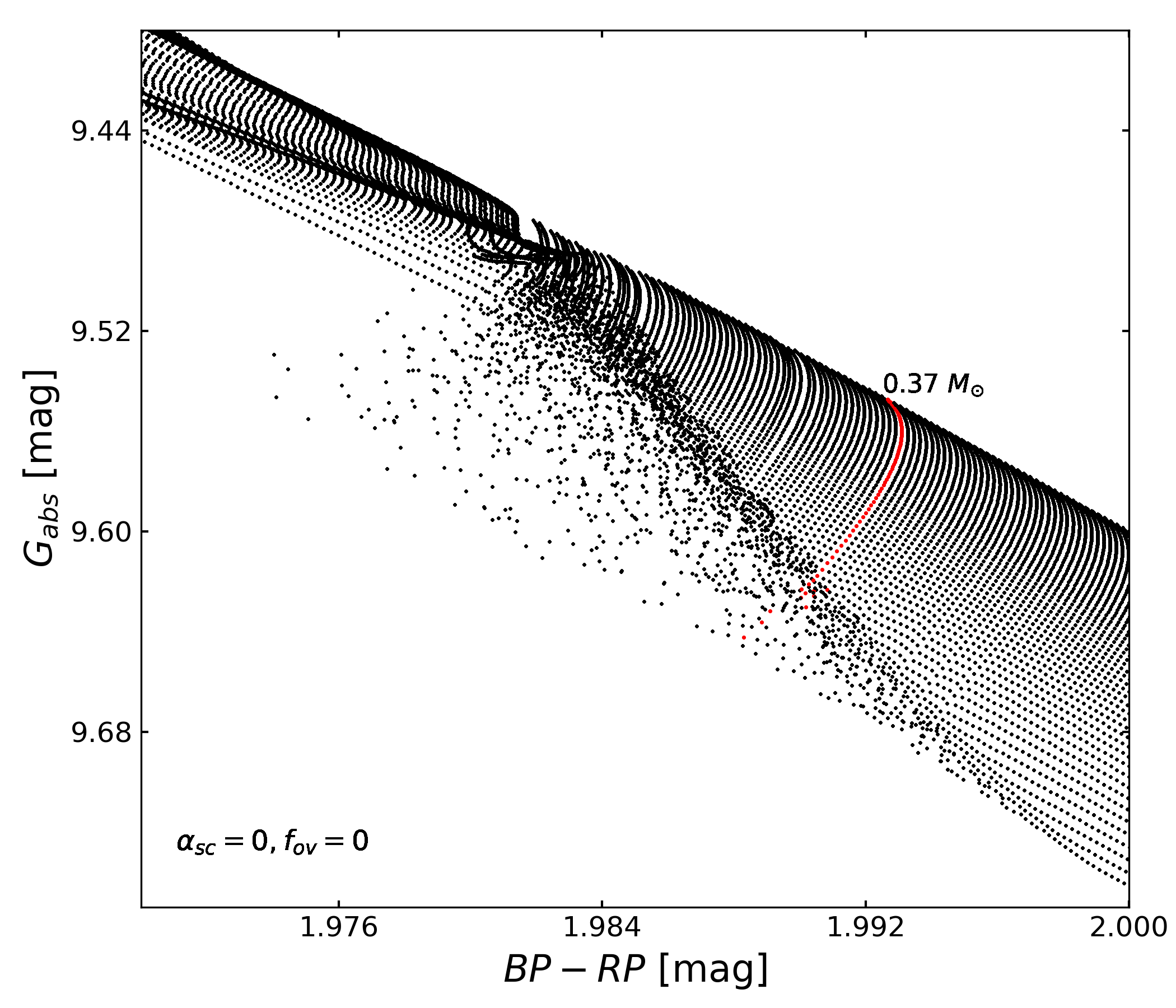}
      \includegraphics[scale = 0.1]{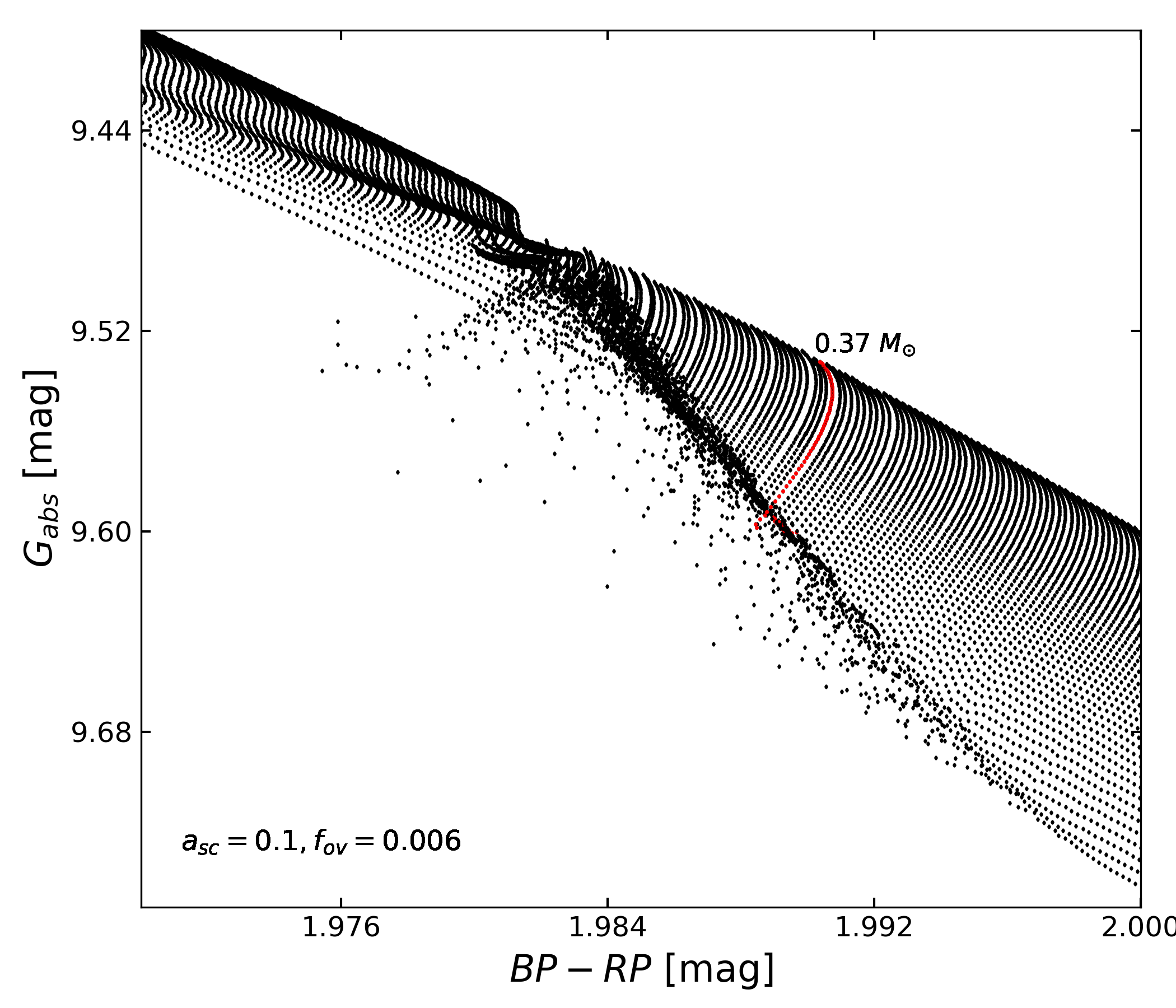}
      \caption{Colour-magnitude diagram showing the main sequence for $Z = 0.02$ across all ages, for no overshooting or semi-convection (\textit{top}), and for $a_{sc} = 0.1$ and $f_{ov} = 0.006$ (\textit{bottom}). The convective kissing instability causes an indent into the blueward edge of the MS. This instability region is more sparsely populated for models with high amounts of overshooting, however it is still present. Marked in red is the 0.37 $M_{\odot}$ model.}
      \label{cmd}
    \end{figure}

    \begin{figure}
    \centering
    \includegraphics[width=\columnwidth]{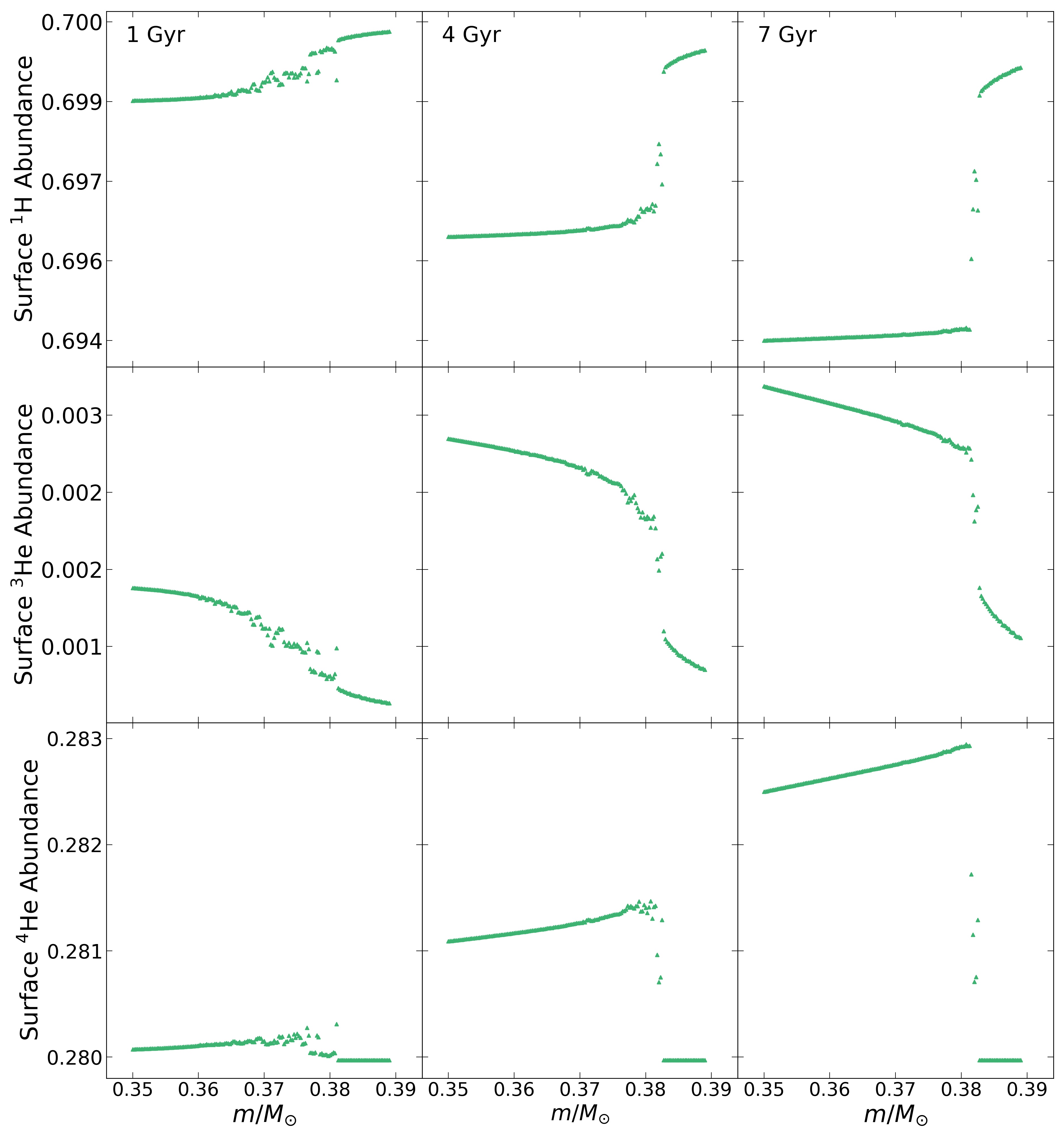}
    \caption{Surface abundances by mass fraction of $^1$H, $^3$He and $^4$He, taken at time intervals of 1, 4 and 7 Gyr, plotted against the mass range for $Z = 0.02$, with no overshooting or semi-convection applied}.
    \label{surfmasstime}
\end{figure}

Figure \ref{overshoot_semi_HR} shows that the loops in the evolutionary tracks of the models in the HR diagrams become much smaller for an increasing amount of overshooting. Whilst overshooting reduces the convective instability, adding semi-convection to the models can retain it. For example, the $f_{ov} = 0.003$ model on the left hand side of Fig. \ref{overshoot_semi} does not fully merge at any point, yet when semi-convection is present, the results become comparable to the model without any overshooting. The radiative region gradually shrinks until the model is fully convective, then a large portion of the material becomes radiative again. With semi-convection, the loops in the HR diagrams are prominent once more, as seen in Fig. \ref{overshoot_semi_HR}. The same is true for a more efficient semi-convection with parameter $\alpha_{sc} = 1$ which is not shown here.

  A comparison of the colour-magnitude diagrams created by the population synthesis for models with no semiconvection or overshooting and for models with $a_{sc} = 0.1$ and $f_{ov} 
= 0.006$ is given in Fig. \ref{cmd}. The indent into the blueward edge of the MS represents the instability region, where the luminosity and temperature of the models fluctuate (as indicated by the loops in Fig. \ref{overshoot_semi_HR}). This instability region is more sparsely populated for the models with high amounts of overshooting, as expected, however it is still distinguishable, meaning that if M-dwarfs were to have high amounts of overshooting, the M-dwarf gap would still be present in the observational CMD. The $0.37 M_{}\odot$ model is marked in Fig. \ref{cmd} which shows a reduction (from top panel to bottom panel) in the area of the instability region over which the model proceeds.

  \subsection{Surface abundances}\label{surf_sec}

An illustration of the changes to the central and surface $^1$H, $^3$He and $^4$He abundances over time is given in Fig. \ref{0.37M}. For a model without overshooting, the rise and drops in the abundances occur when the model becomes fully convective and mixes the stellar material throughout the model. When overshooting is applied (Fig. \ref{3he_crit_abund}) the material is mixed without the model being fully convective. In

      \begin{landscape}
      \begin{figure}
      \hspace*{-5pt}
      \includegraphics[scale=0.16]{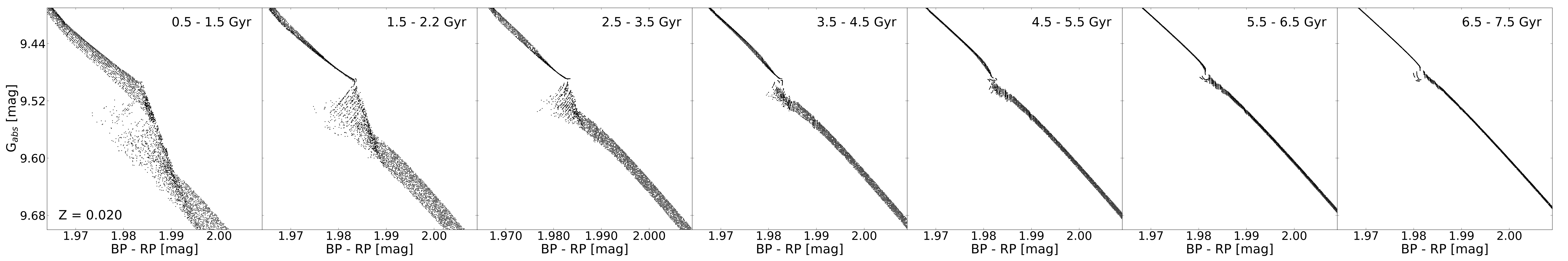}
      \caption{Composite Galactic-field populations grouped into age bins spanning 1 Gyr at increasing population age, showing that the width of the MS decreases with time. Here no overshooting or semi-convection is used.} 
      \label{cmd_zoom}
    \end{figure}

    \begin{figure}
     \hspace*{-5pt}
      \includegraphics[scale=0.16]{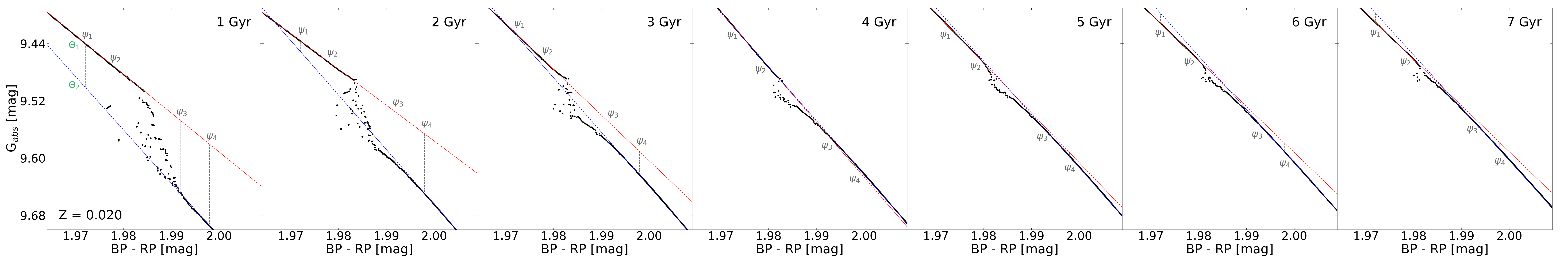}
      
      \hspace*{-5pt}
    \includegraphics[scale=0.16]{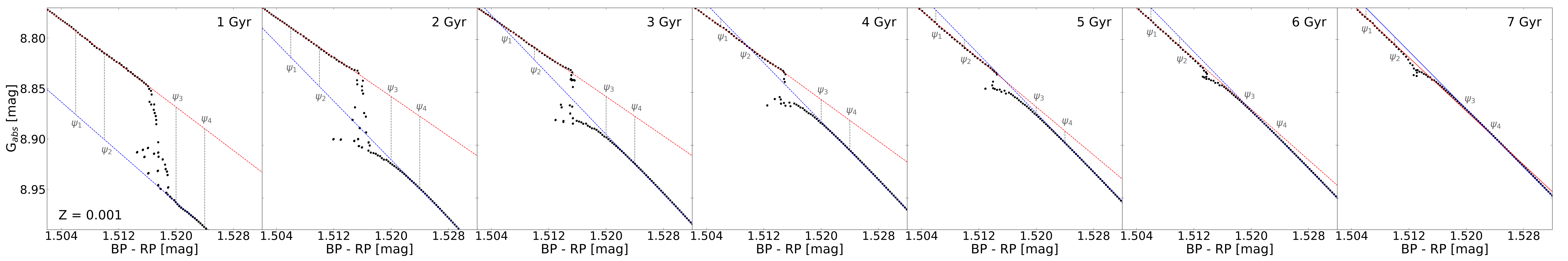}
        \caption{A set of solar metallicity stars (\textit{top}) and with $Z = 0.001$ (\textit{bottom}) are given where the stars have the same age and no overshooting or semi-convection is used. The changes in width can no longer be seen without any age distribution, but the parallel offset of the MS and the kink due to the convective kissing instability remain. The offsets, $\Psi_i$, and the relative angle, $\theta = \theta_1 - \theta_2$, of the upper and lower parts of the MS are detailed in Fig. \ref{offset}.}
        \label{ms_psi}
    \end{figure}
\end{landscape}
 
\noindent either case, the surface abundance of helium continuously rises (and hydrogen decreases) until it is equal to the central abundance.

If we view the surface abundances across the mass range, we can see the difference between the models which are fully convective and those that are not. Figure \ref{surfmasstime} shows the surface abundances of $^1$H, $^3$He and $^4$He by mass fraction plotted against the mass range, given at time intervals of 1, 4 and 7 Gyr. As the models age, a drop forms within the model sets, which represents the convective boundary. The models with masses higher than the drop are those with radiative cores and convective envelopes, and as such the surface abundances of these do not vary significantly over time. The models with masses lower than the drop are fully convective, and so the surface abundances of $^3$He and $^4$He steadily increase with time as these isotopes that are being produced in the core are carried outward to the surface by convection. Conversely, the surface abundance of $^1$H decreases over time for the fully convective models, as $^1$H is mixed downward re-fuelling the core.

\begin{figure}
    \centering
    \includegraphics[scale=0.06]{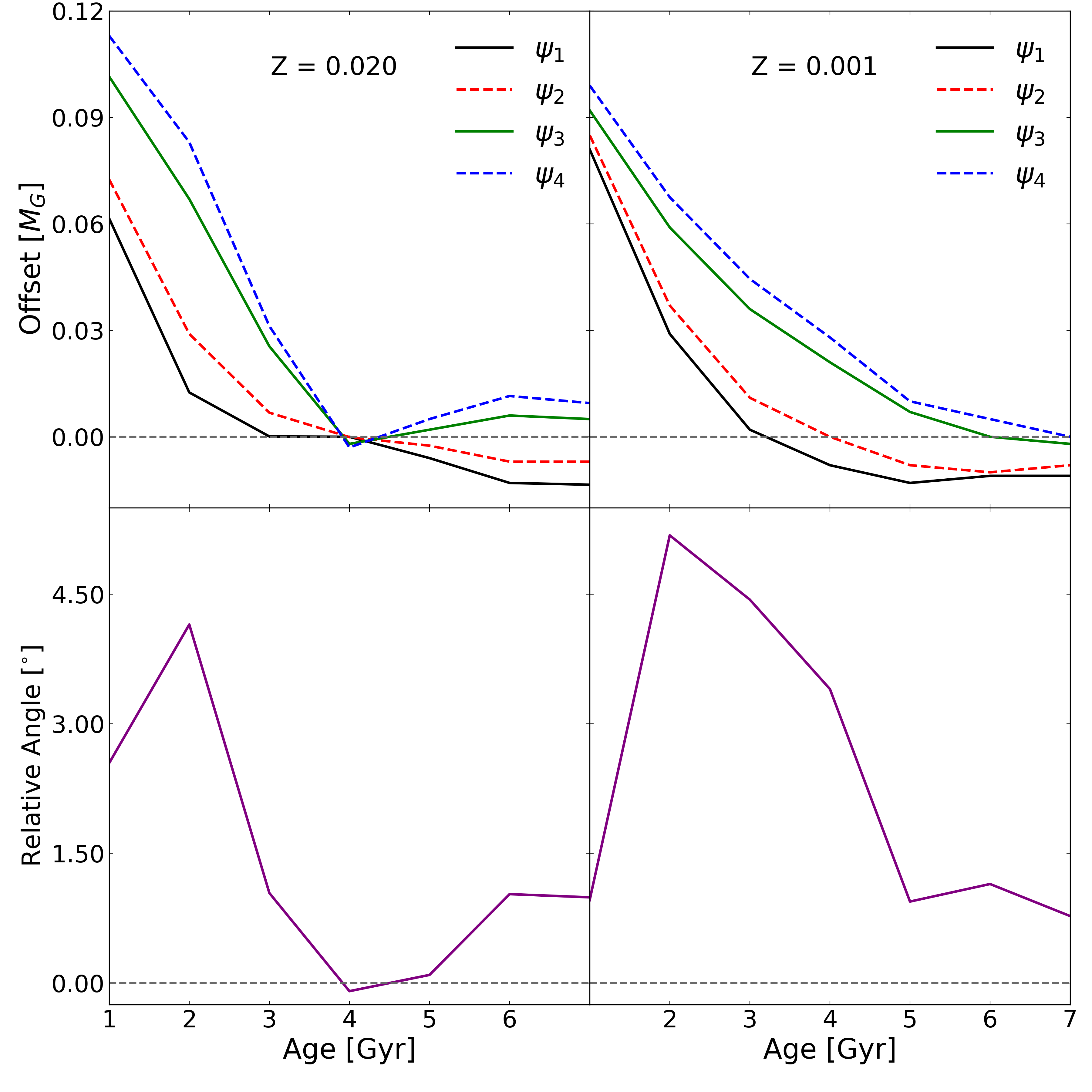}
    \caption{The offsets $\Psi_i$ (\textit{top}),  and the relative angle (\textit{bottom}) between the upper and lower parts of the MS shown in Fig. \ref{ms_psi}.}
    \label{offset}
\end{figure}

The drop illustrated in Fig. \ref{surfmasstime} moves to higher masses with time. This was seen in \citet{mansfield} as the lower mass models undergo the instability at an earlier point in their lifetimes than the higher mass models, and once the instability ceases, the models remain fully convective. This drop would still be present in the absence of the convective kissing instability, however it would likely be a sudden discontinuity which would not change with time. 

 The average relative changes in surface abundance by the end of the model lifetime is approximately a 1.0\% decrease for $^1$H and a 1.2 \% increase of $^4$He and as such are unlikely to be detected over observational noise. The surface abundance of $^3$He has quadrupled although it still remains a small fraction of the overall stellar material. 

Similarly to hydrogen and helium, once the lower mass models have undergone the convective kissing instability and remain fully convective, the surface $^{12}$C and $^{16}$O values decrease, and $^{14}$N increases, however these abundances change by less than 0.2\% during the total model evolution. As the differences are so small, they are not shown here. 
  
  \subsection{The width of the MS: the Solar neighbourhood}\label{cmd_sec}
  
  As seen in Sect. \ref{overshoot_sec}, the convective kissing instability presents as an indent into the blueward edge of the MS (Fig. \ref{cmd}) in the colour-magnitude diagram from our population synthesis. The width of the MS varies significantly across this region. Figure \ref{cmd_zoom} shows the CMD at increasing population age, representing composite Galactic-field populations, where the overall width of the MS decreases over time. Additionally, the width of the MS is larger for stars with masses lower than the instability than for masses higher, and with time the difference between these two widths also becomes smaller.

   \begin{table}
       \centering
       \caption{The parallel offsets and relative angle between the upper and lower parts of the MS as shown in Fig. \ref{ms_psi}, for $Z = 0.02$}
       \begin{tabular}{crrrrr}
Age & $\Psi_1$\ \ \ & $\Psi_2$\ \ \ & $\Psi_3$\ \ \ & $\Psi_4$\ \ \ & $\theta$\ \ \ \ \\

[Gyr] &[$M_G$]\ &[$M_G$]\ &[$M_G$]\ &[$M_G$]\ &[$^{\circ}$]\ \ \\ \hline\hline
 1 & 0.062 & 0.073 & 0.102 & 0.113 & 2.551\\
 2 & 0.013 & 0.029 & 0.067 & 0.083 & 4.155\\
 3 & 0.001 & 0.007 & 0.026 & 0.031 & 1.042\\
 4 & 0.000 & -0.001 & -0.002 & -0.003 & -0.094\\
 5 & -0.006 & -0.003 & 0.002 & 0.005 & 0.094\\
 6 & -0.013 & -0.007 & 0.006 & 0.115 & 1.029\\
 7 & -0.014 & -0.007 & 0.005 & 0.010 & 0.992\\
\hline
     
       \end{tabular}
       \label{offset_table}
   \end{table}

      \begin{table}
       \centering
       \caption{Same as Table \ref{offset_table} but for $Z = 0.001$}
       \begin{tabular}{crrrrc}
Age & $\Psi_1$\ \ \ & $\Psi_2$\ \ \ & $\Psi_3$\ \ \ & $\Psi_4$\ \ \ & $\theta$\\

[Gyr] &[$M_G$]\ &[$M_G$]\ &[$M_G$]\ &[$M_G$]\ &[$^{\circ}$]\ \ \\ 
\hline\hline
1 & 0.081 & 0.085 & 0.092 & 0.099 & 0.960\\
2 & 0.029 & 0.037 & 0.059 & 0.068 & 5.181\\
3 & 0.002 & 0.011 & 0.036 & 0.045 & 4.437\\
4 & -0.008 & 0.000 & 0.021 & 0.028 & 3.403\\
5 & -0.013 & -0.008 & 0.007 & 0.010 & 0.943\\
6 & -0.011 & -0.010 & 0.000 & 0.005 & 1.146\\
7 & -0.011 & -0.008 & -0.002 & 0.000 & 0.775\\
\hline
     
       \end{tabular}
       \label{offset_table2}
   \end{table}
   
   \subsection{Age dating: star clusters}

   Figure \ref{ms_psi} gives a set of CMDs for $Z = 0.020$ and a lower metallicity of $Z = 0.001$ with stars of the same age, representing a Solar-neighbourhood open cluster and a Galactic halo globular cluster respectively. Although the absence of an age spread does not show the development in MS width in this case, we can still see the change in parallel offset between the upper and lower parts of the MS, as well as the large kink due to the convective kissing instability. The offsets $\Psi_i$ measured at four points ($BP\ -\ RP = 1.972, 1.978, 1.992$ and $1.998$ for $Z = 0.020$, and $BP\ -\ RP = 1.506, 1.510, 1.520$ and $1.524$ for $Z = 0.001$) are quantified in Tables \ref{offset_table} \& \ref{offset_table2} and illustrated in Fig. \ref{offset}, along with the relative angle from vertical, $\theta = \theta_1 - \theta_2$, between the upper and lower MS. For $Z = 0.020 $, the directions of the MS at masses higher (upper MS) and lower (lower MS) than the convective kissing instability are considerably divergent by 2 Gyr and then become more in line with each other as the models evolve, except for at 4 Gyr when the sign of the relative angle becomes reversed. Similarly for $Z = 0.001$, the MS is roughly parallel at the beginning of the evolution, the upper and lower parts of the MS diverge by 2 Gyr, then become increasingly more aligned until 7 Gyr when they are nearly parallel again. By measuring the parallel offset and the relative angle between the upper and lower MS, the age of a star cluster could be estimated. 

\subsection{Age dating: single M-dwarf stars}

\begin{figure}
       \centering
       \includegraphics[scale = 0.12]{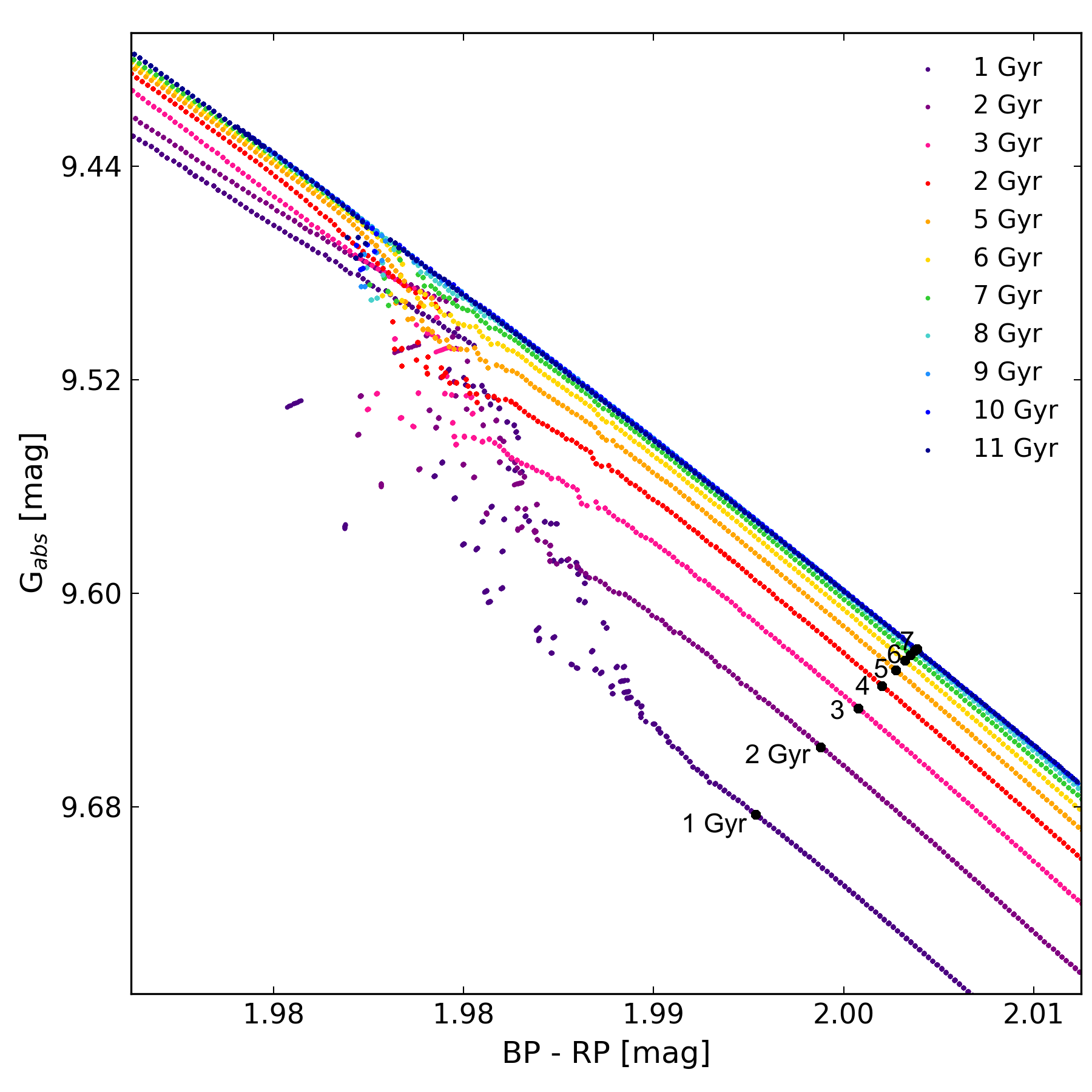}
       \caption{A CMD showing the MS of M-dwarf stars at different snapshots in time, without overshooting or semi-convection applied and a metallicity of $Z = 0.02$. Marked on the diagram is a single model at different ages. If the position in the CMD of a star can be accurately known, along with its metallicity, then the age of the star could be determined, until around 7 Gyr.}
       \label{cmd_gaia_one}
   \end{figure}

   A method for potential age dating of an M-dwarf star with known distance, metallicity and mass is by the star's position in the CMD, if the mass is lower than the convective kissing instability. This is illustrated in Fig. \ref{cmd_gaia_one} where the time-dependent location in the CMD of a single mass model is given. As the model evolves, it moves redward and towards higher luminosities, but only up to a point, meaning that this method could only be used to estimate the age of a star up until around 7 Gyr.

  \subsection{Mass-magnitude relation and the luminosity function}\label{lum_sec}

\begin{figure}
      \centering
      \includegraphics[scale=0.1]{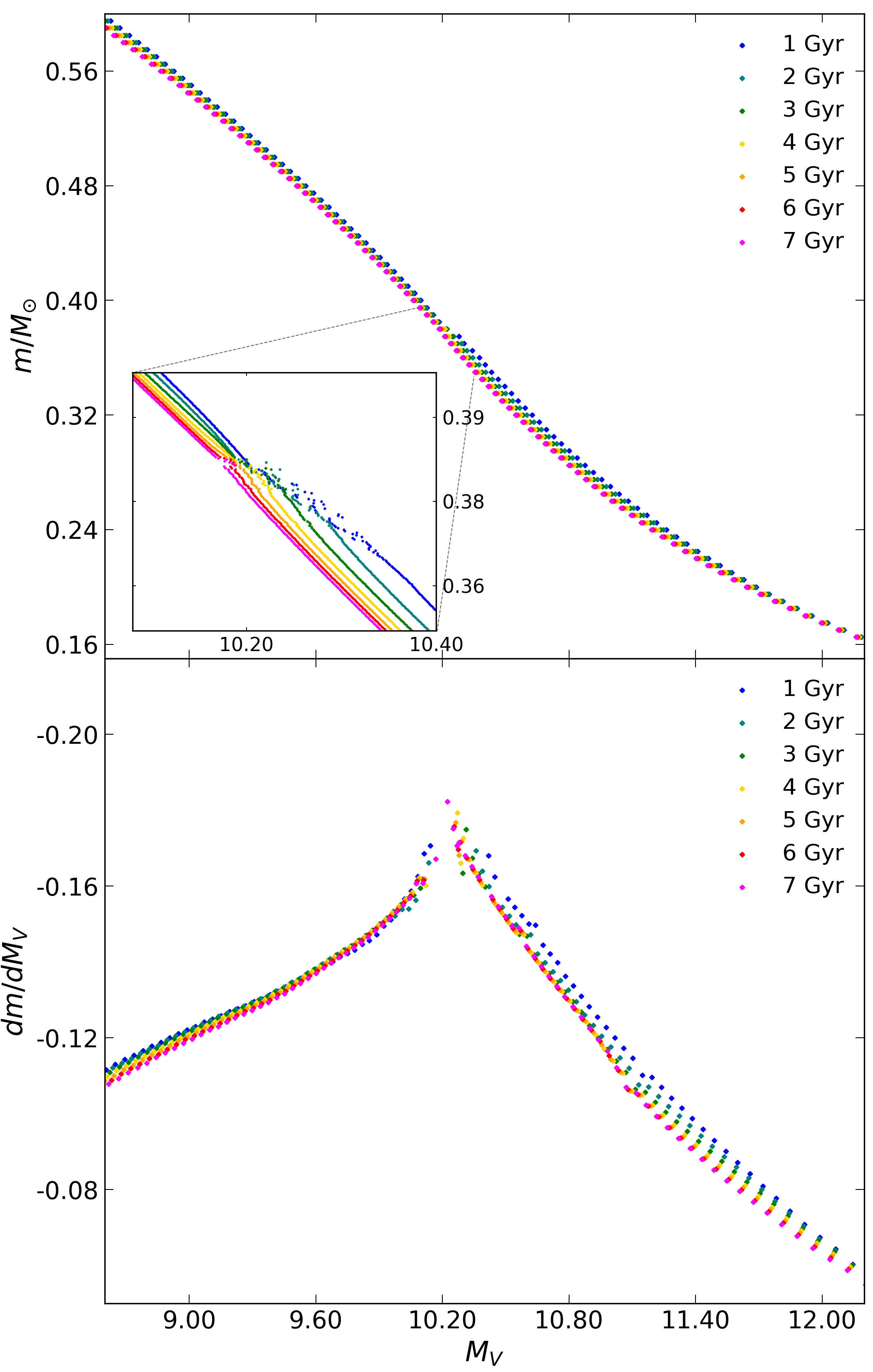}
      \caption{Mass -- $M_{V}$ relation (\textit{top}) and its derivative (\textit{bottom}) for solar metallicity and no overshooting or semi-convection applied, showing a slight dependence in time for the models with masses lower than the convective kissing instability. Inset is a zoomed in region of the instability, showing the discontinuity.}
      \label{dmdMV}
  \end{figure}

Figure \ref{dmdMV} shows the mass-magnitude relation and its derivative, $dm/dM_V$, with the convective kissing instability shown in the inset plot. The gap in the peak of the derivative at $M_V = 10.20$ is due to the discontinuity being indifferentiable. Just as the width of the MS in the CMD is wider for models with masses below the instability when viewed over time, the mass-magnitude relation and its derivative also exhibits a time dependence for these masses. 
  
\begin{figure}
      \centering
      \includegraphics[width=\columnwidth]{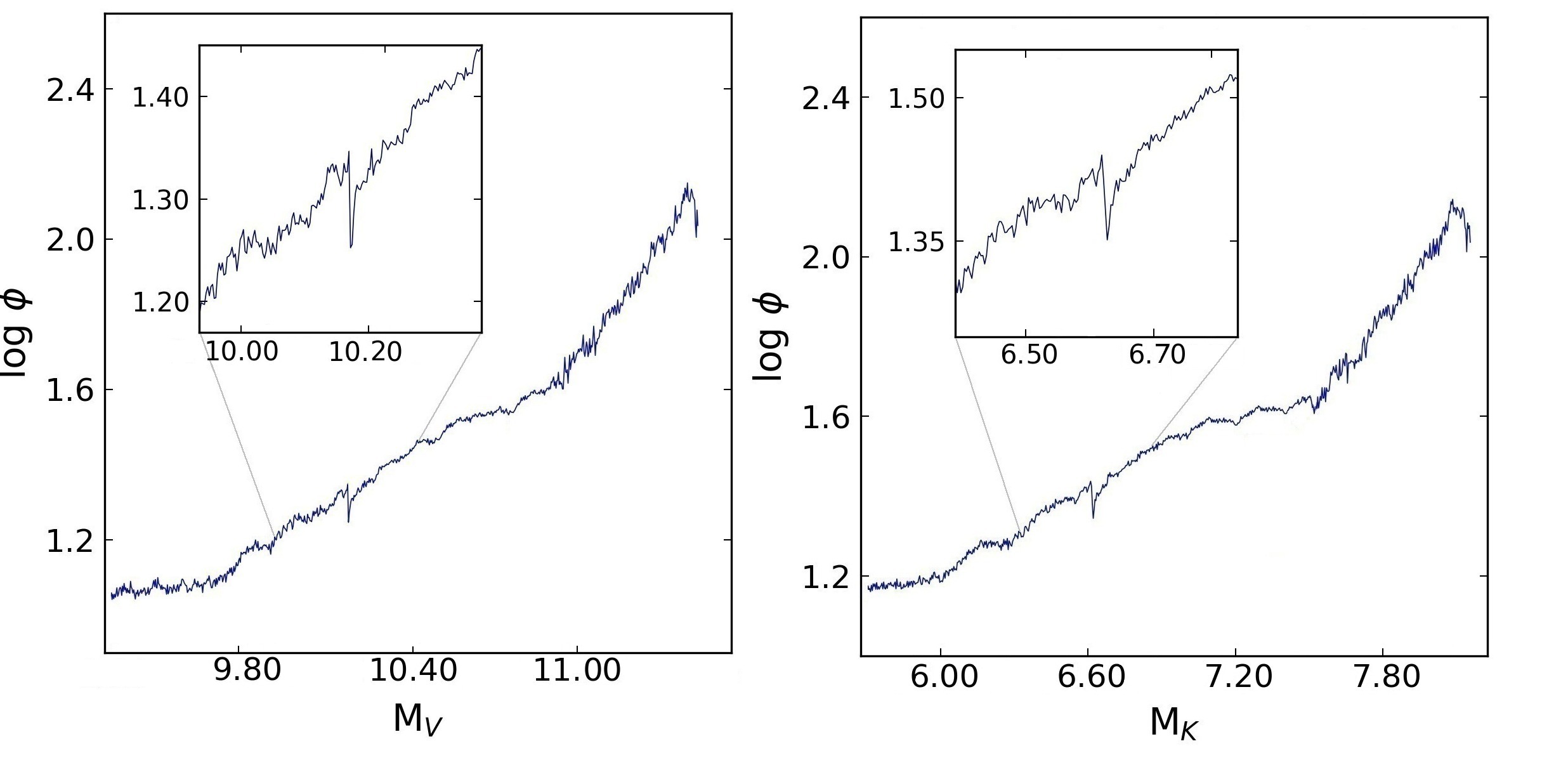}
      \caption{Stellar luminosity function in the $V$ and $K$ wavebands for $Z = 0.02$ and all ages and no overshooting or semi-convection applied}, showing a peak and dip due to the convective kissing instability.
      \label{lf}
  \end{figure}

  Figure \ref{lf} shows the stellar luminosity function in
  different wavebands. At the magnitude where the convective kissing instability occurs, there is a small peak and dip in the number of stars at $M_V = 10.20$ and $M_K = 6.63$. The inset plots give a closer view of this feature. The dip corresponds to the indent into the blueward edge of the MS, and the small peak that is brighter in magnitude indicates a slight over-density which can be seen in Fig. \ref{cmd} as the wave-like overlapping of models just above the gap. This peak is likely to correlate to the same slight over-abundance of stars seen close to the gap in observations \citep{ Jao_2020}. 

\section{Discussion}\label{discus}

The convective kissing instability is reduced with increasing amounts of overshooting. The loops in the evolutionary tracks of the models in the HR diagram also become suppressed for increasing overshooting. This brings into question whether M-dwarf stars undergoing the convective kissing instability with overshooting can still be responsible for the M-dwarf gap. The CMD created from the population synthesis (Fig. \ref{cmd}) using a relatively high amount of overshooting indicates that there are fewer models within the instability region, yet it does remain. If the amount of overshooting is low, and semi-convection is also present, then the loops in the HR diagram are sustained (Fig. \ref{overshoot_semi_HR}). Therefore in order to reproduce sufficient fluctuations in luminosity and temperature to cause the observed M-dwarf gap, stars at this mass range may have overshooting, with less overshooting being preferable (we suggest an upper limit of $f_{ov,max} = 0.004$), and semi-convection should also be present. 

The indent that lies in the blueward edge of the MS is more pronounced for the younger models, similar to the larger kinks found in lower-age isochrones given by \citealt{baraffe} (their fig. 6). This is due to the models settling into a fully convective state after several gigayear \citep{mansfield} and the cessation of the fluctuations in luminosity and temperature. From our models we see that there is a reduction in the width of the MS over time for composite Galactic-field populations, and a difference in width of the MS at masses lower than the instability compared with masses higher (Figs. \ref{cmd_zoom}). For star clusters, which can be assumed to have members of all the same age, the parallel offset and relative angle between the upper and lower MS changes with time (Figs. \ref{ms_psi} \& \ref{offset}). This is detected in a recently observed CMD of the Hyades cluster, which finds a difference in MS direction for stars higher and lower than $0.35M_{\odot}$ (for a super-solar metallicity of $\textnormal{[Fe/H]} = +0.25$, \citealt{Brandner}, their fig. 1). These findings then have the potential to be used as an age estimator of stellar populations. The width of the MS may also be used to estimate the age of single M-dwarf stars if the precise metallicity and mass are known (Fig. \ref{cmd_gaia_one}).

The stellar luminosity function shows a small feature located at around $M_V \approx 10.20$ (Fig. \ref{lf}), which is at a slightly brighter magnitude than the prediction by  \citet{jao} of finding a gap in the stellar luminosity function at around $M_V \approx 10.5 - 11.5$. A dip in the luminosity function at $M_K = 6.71$ is also predicted by \citet{macdonald}, which is in agreement to the dip seen in Fig. \ref{lf} at $M_K \approx 6.63$. \citealt{macdonald} propose an alternative theory to explain the M-dwarf gap in which convective mixing is treated as a diffusive process, resulting in their models undergoing a single merger event rather than the convective kissing instability. However, they report that the central $^3$He abundance is increased during this merger from the mixing of $^3$He from the envelope, yet we find the opposite relation in our models. The central $^3$He abundance is reduced when the model becomes fully convective and material is mixed throughout the star (Fig. \ref{0.37M}).

The convective kissing instability effects the surface abundance values of the elements $^{1}$H, $^{3}$He, $^{4}$He, $^{12}$C, $^{14}$N and $^{16}$O through the core and envelope merger events when the models are fully convective and material is transported throughout the model. The instability effects higher mass models over time and the lower masses settle into a fully convective state. This can be seen as the relative changes in abundance values move to higher masses over time. If the abundance values of a star could be measured and the metallicity is also known, this could be used to estimate the age and mass of the star. However, the relative changes for all elements and isotopes studied are very small and likely will not be larger than observational noise, as well as being within the normal variations of abundances due to stars being formed at different locations and times. Thus, whilst interesting from the evolutionary model standpoint, the predicted changes in surface abundances will be challenging to  detect in observations.

\section{Summary}\label{sum}

The amplitude and intensity of the convective kissing instability is reduced with increasing amounts of overshooting but sustained when semi-convection is also present. This effect is seen to modify the loops in the evolutionary tracks in the HR diagram.

The surface abundance values of M-dwarf models are altered by moments of full convection as well as once the models settle into a fully convective state. However the relative changes are small and unlikely to be detected significantly enough to be separated from observational noise or natural variation.

The merging of the convective core and convective envelope produces a discontinuity in the mass-magnitude relation and we are able to take the magnitudes from the models at multiple moments in time to create the stellar luminosity function. This synthetic population of stars shows the M-dwarf gap as an indent into the blueward edge of the MS as well as an overabundance of stars at slightly brighter magnitudes where the models overlap. This results in a feature consisting of a small peak and dip in the luminosity function near $M_V \approx 10.20$.

The width of the MS for a Galactic-field composite population decreases over time, as well as the difference in width of the MS from stars with masses higher and lower than the instability. The parallel offset and relative angle between the upper and lower MS for a single-age star cluster changes with time. This may be used to estimate the ages of stellar populations. Due to the large width of the lower MS, the position of a star in the CMD with known metallicity and mass may also be used to estimate the age of the star if the mass is close to but lower than the convective kissing instability. Furthermore, the width of the lower MS also correlates to a time dependence in the mass-magnitude relation and its derivative. 

Finally, a consideration for future study is to investigate different mixing lengths and rotation. Rotation can drive strong mixing of material within stars, and M-dwarfs have been observed to rotate quickly at the young ages, and thus an investigation into rotation would also be beneficial for further study.

\section*{Data availability}

The datasets were derived from MESA which is in the public domain:  \url{https://docs.mesastar.org/}

\section*{Acknowledgements} 

We thank the referee for the detailed and thoughtful comments.


\bibliographystyle{mnras}
\bibliography{MDwarfGap_ed.bib}

\end{document}